\documentstyle[sprocl]{article}

\def\be{\begin{equation}}
\def\ee{\end{equation}}
\def\bea{\begin{eqnarray}}
\def\eea{\end{eqnarray}}
\newcommand{\g}{{\sl g}}
\newcommand{\cD}{{\cal D}}
\newcommand{\cG}{{\cal G}}
\newcommand{\cI}{{\cal I}}
\newcommand{\cO}{{\cal O}}
\newcommand{\cN}{{\cal N}}
\newcommand{\cK}{{\cal K}}
\newcommand{\cQ}{{\cal Q}}

\newcommand{\OO}{\mathop{\otimes}}
\newcommand{\ft}[2]{{\textstyle\frac{#1}{#2}}}

\def\1{\hbox{{1}\kern-.25em\hbox{l}}}

\begin{document}

%%%%%%%%%%%%%%%%%%%%%%%%%%%%%%%%%%%%%%%%%%%%%%%%%%%%%%%%%%%%%%%%%%%%%%%%
\begin{titlepage}

\centerline{\large \bf NLO Exclusive Evolution Kernels.}

\vspace{10mm}

\centerline{\bf A.V. Belitsky$^a$, A. Freund$^b$, D. M\"uller$^c$}

\vspace{10mm}

\centerline{\it $^a$C.N. Yang Institute for Theoretical Physics}
\centerline{\it     State University of New York at Stony Brook}
\centerline{\it     NY 11794-3800, Stony Brook, USA}

\vspace{3mm}

\centerline{\it $^b$INFN, Sezione di Firenze, Largo E. Fermi 2}
\centerline{\it     50125, Firenze, Italy}

\vspace{3mm}

\centerline{\it $^c$Institut f\"ur Theoretische Physik,
                 Universit\"at Regensburg}
\centerline{\it     D-93040 Regensburg, Germany}

\vspace{30mm}

\centerline{\bf Abstract}

\vspace{0.8cm}

We outline a formalism used for a construction of two-loop flavor singlet
exclusive evolution kernels in the $\overline{\rm MS}$ scheme. The approach
is based on the known pattern of conformal symmetry breaking in
$\overline{\rm MS}$ as well as constraints arising from the superconformal
algebra of the $\cN = 1$ super Yang-Mills theory.

\vspace{30mm}

\centerline{\it Talk given at the}
\centerline{\it 8th International Workshop on Deep Inelastic
                Scattering and QCD}
\centerline{\it Liverpool, April 25-30, 2000}

\end{titlepage}
%%%%%%%%%%%%%%%%%%%%%%%%%%%%%%%%%%%%%%%%%%%%%%%%%%%%%%%%%%%%%%%%%%%%%%%%

\title{NLO Exclusive Evolution Kernels}

\author{A. V. Belitsky$^a$, A. Freund$^b$, D. M\"uller$^c$}

\address{$^a$C.N. Yang ITP, SUNY at Stony Brook\\
             NY 11794-3800, Stony Brook, USA}
\address{$^b$INFN, Sezione di Firenze, Largo E. Fermi 2\\
             50125, Firenze, Italy}
\address{$^c$Institut f\"ur Theoretische Physik, Universit\"at Regensburg\\
             D-93040 Regensburg, Germany}

\maketitle\abstracts{We outline a formalism used for a construction of
two-loop flavor singlet exclusive evolution kernels in the $\overline{\rm
MS}$ scheme. The approach is based on the known pattern of conformal
symmetry breaking in $\overline{\rm MS}$ as well as constraints arising
from the superconformal algebra of the $\cN = 1$ super Yang-Mills theory.}

%%%%%%%%%%%%%%%%%%%%%%%%%%%%%%%%%%%%%%%%%%%%%%%%%%%%%%%%%%%%%%%%%%%%%%%%
\section{$Q^2$ evolution of SPD}
%%%%%%%%%%%%%%%%%%%%%%%%%%%%%%%%%%%%%%%%%%%%%%%%%%%%%%%%%%%%%%%%%%%%%%%%

Exclusive processes provide an indispensable information for a construction
of a unique picture of hadron wave functions, ${\mit \Psi}$. Its lowest Fock
components (integrated over different transverse momentum configurations of
partons) go under the name of distribution amplitudes, $\phi (x)$. Being a
fundamental characteristic, ${\mit\Psi}$ defines all other inclusive and
exclusive observables. A product of a wave function and a complex conjugate
with fixed transverse momentum
\be
\phi \left( x, \eta, \Delta_\perp \right) \sim \int d^2 k_\perp
{\mit\Psi}^\ast
\left( \ft{x + \eta}{2}, k_\perp + \ft{\Delta_\perp}{2} \right)
{\mit\Psi}
\left( \ft{x - \eta}{2}, k_\perp - \ft{\Delta_\perp}{2} \right) ,
\ee
define a correlation function called skewed parton distribution (SPD). It
generalizes its predecessor, --- conventional inclusive density well known
from DIS, to non-zero values of skewedness $\eta$ and $\Delta_\perp$. A
peculiar feature of the SPDs is that they have a very different behaviour
depending on the kinematical regime, i.e.\ an interplay of $x$ and $\eta$.
Depending on the difference in the momentum fractions between the left- and
right-hand-side of the parton ladder the SPDs behave like a regular parton
distribution or like a distribution amplitude. Particular Mellin moments
w.r.t.\ momentum fraction $x$ give hadron (and real Compton scattering) form
factors and angular momenta of constituents.

In QCD the leading twist SPD is defined as a Fourier transform to the
momentum fraction space of a light-ray operator constructed from
$\varphi$-parton fields and sandwiched between hadronic states non-diagonal
in momenta, schematically given by ($\Delta = p' - p$)
\be
\label{DefSPD}
\phi (x, \eta, \Delta_\perp | Q)
= \frac{1}{2\pi} \int d z_- e^{i x z_-}
\langle h (p') |
\left. \varphi^\dagger ( - z_-/2 ) \varphi ( z_-/2 ) \right|_{Q}
| h (p) \rangle
\ee
The logarithmic $Q$-scale dependence of $\phi$ arises due to a light-like
separation of partons and is governed by a renormalization group equation.
The generalized skewed kinematics for corresponding perturbative evolution
kernels can unambiguously be restored from the conventional exclusive one
$\eta = 1$.
\be
\label{ER-BLequation}
\frac{d}{d \ln Q^2} \mbox{\boldmath$\phi$} (x | Q)
= \mbox{\boldmath$V$} \left(x, y | \alpha_s(Q) \right)
\mathop{\otimes} \mbox{\boldmath$\phi$} (y | Q) ,
\ee
where $\tau_1 \OO \tau_2 (x,y) \equiv \int_0^1 dz\, \tau_1(x,z) \tau_2
(z, y)$ defines the exclusive convolution and $\mbox{\boldmath$\phi$}
= \left(\phi^Q, \phi^G\right)$ is the vector of the quark and gluon
distributions and $\mbox{\boldmath$V$}$ is a matrix of evolution kernels.
Thanks to conformal invariance of classical QCD Lagrangian the leading
order kernels having the structure $\mbox{\boldmath$V$}^{(0)} (x, y) =
\theta (y - x) \mbox{\boldmath$f$} (x, y) \pm \theta (x - y)
\mbox{\boldmath$f$} (\bar x, \bar y)$ can be diagonalized in the basis
spanned by Gegenbauer polynomials $C_j^\nu (x) \OO \mbox{\boldmath$V$}^{(0)}
(x, y) = \mbox{\boldmath$\gamma$}_j^{(0)} C_j^\nu (y)$ with forward
anomalous dimensions (ADs) $\mbox{\boldmath$\gamma$}_j^{(0)}$. Beyond
this level conformal symmetry is violated by quantum corrections and
a diagonal AD matrix $\mbox{\boldmath$\gamma$}_j$ gets promoted to a
triangular one $\mbox{\boldmath$\gamma$}_{jk}$, $k \leq j$. Thus $\mbox{
\boldmath$V$} = \mbox{\boldmath$V$}^{\rm D} + \mbox{\boldmath$V$}^{\rm ND}$
with $\mbox{\boldmath$V$}^{\rm ND} \propto \cO (\alpha_s^2)$. An efficient
formalisms to tackle the problem which eludes explicit multi-loop
exercise and is based on the use of special conformal anomalies which
produce the non-diagonal part, $k < j$, of $\mbox{\boldmath $\gamma$}_{jk}$,
converted into exclusive kernels $\mbox{\boldmath$V$}^{\rm ND}$; and
relations resulting from $\cN = 1$ SUSY Ward identities which connect
diagonal part of the kernels, $\mbox{\boldmath$V$}^{\rm D}$, and allows
to reconstruct all channels from a given $QQ$ sector deduced by explicit
evaluation of two-loop graphs.

%%%%%%%%%%%%%%%%%%%%%%%%%%%%%%%%%%%%%%%%%%%%%%%%%%%%%%%%%%%%%%%%%%%%%%%%
\section{Using conformal symmetry}
%%%%%%%%%%%%%%%%%%%%%%%%%%%%%%%%%%%%%%%%%%%%%%%%%%%%%%%%%%%%%%%%%%%%%%%%

Conformal operators which are Gegenbauer moments of $\phi$,
$C_j^\nu (x) \OO \phi (x) \sim \langle h' | \cO_{jj} | h \rangle$, build an
infinite dimensional irreps of the collinear conformal algebra $so(2,1)$.
Conformal Ward identities derived for the Green function with conformal
operator insertion $\cG \equiv \langle \cO_{jk} \prod_i \varphi_i \rangle$
in the regularized QCD allows, by means of algebra of dilatation $\cD$ and
special conformal transformation $\cK$, to prove a matrix constraint for
ADs $\mbox{\boldmath{$\gamma$}}$ and special conformal anomaly
$\mbox{\boldmath{$\gamma$}}^c$
\be
\label{conf-constr-KD-1}
\left[ \cD , \cK_- \right]_- = i \cK_-
\qquad\Rightarrow\qquad
\left[ \mbox{\boldmath{$a$}}
+ \mbox{\boldmath{$\gamma$}}^c
+ 2 \ft{\beta}{\g} \, \mbox{\boldmath{$b$}}  ,
\mbox{\boldmath{$\gamma$}} \right]_- = 0 ,
\ee
with $\alpha_s$-independent matrices $\mbox{\boldmath{$a$}}$ and $\mbox{
\boldmath{$b$}}$ and QCD beta function $\beta = \ft{\alpha_s}{4\pi} \beta_0
+ \cdots$. The solution of the above equation with available one-loop
conformal anomalies $\mbox{\boldmath{$\gamma$}}^c$ implies the following
form of the nondiagonal part of the NLO kernel
\be
\label{def-ND-kernel}
\mbox{\boldmath$V$}^{{\rm ND}(1)} (x, y)
= - ( \cI - \cD )\,
\left\{
\mbox{\boldmath$\dot V$} \OO
\left(
\mbox{\boldmath$V$}^{(0)} + \ft{\beta_0}{2}\, \1
\right)
+
\left[
\mbox{\boldmath$g$} \OO , \mbox{\boldmath$V$}^{(0)}
\right]_-
\right\} (x, y) ,
\ee
where $( {\cal I} - {\cal D} )$ projects out the diagonal part $\mbox{
\boldmath{$\gamma$}}^{(1)}_{jj}$. Here $\mbox{\boldmath$\dot V$}$ is given
mostly by a logarithmic modification of LO kernels $\mbox{\boldmath$f$}
\to \mbox{\boldmath$f$} \ln\ft{x}{y}$ plus an addendum, while $\mbox{
\boldmath$g$}$ is a kernel whose conformal moments are proportional to a
$\mbox{\boldmath{$w$}}$ part of $\mbox{\boldmath{$\gamma$}}^c
= -\mbox{\boldmath{$b$}} \mbox{\boldmath{$\gamma$}}^{(0)} +
\mbox{\boldmath{$w$}}$.

%%%%%%%%%%%%%%%%%%%%%%%%%%%%%%%%%%%%%%%%%%%%%%%%%%%%%%%%%%%%%%%%%%%%%%%%
\section{Using $\cN = 1$ SUSY}
%%%%%%%%%%%%%%%%%%%%%%%%%%%%%%%%%%%%%%%%%%%%%%%%%%%%%%%%%%%%%%%%%%%%%%%%

The last problem is to find $\mbox{\boldmath$V$}^{\rm D}$. Although
it seems straightforward to solve, since the Gegenbauer moments
$\mbox{\boldmath$V$}^{\rm D}_{jk} = \delta_{jk} \mbox{\boldmath$\gamma$}_j$
coincide with forward ADs calculated to NLO presently, practical inversion
is extremely hard to handle. The main difficulty being kernels stemming from
crossed-ladder type diagrams which we called \mbox{\boldmath$G$}-functions.
Since the conformal symmetry breaking part has been previously fixed, we
can assume conformal covariance for the ADs. If one puts (Majorana) quarks
into adjoint representation of $SU (N_c)$, the classical ``QCD'' Lagrangian
enjoys $\cN = 1$ SUSY. In perturbative calculations (with SUSY preserving
regularization) this simply means the following identification of Casimir
operators: $C_F=2T_F=C_A\equiv N_c$. From the commutator of the dilatation
and translational SUSY generators $[\cQ, \cD]_- = \ft{i}{2} \cQ$ applied
to the Green functions $\cG$ one finds six constraints for eight
$\mbox{\boldmath$G$}$-functions for even and odd parity sectors. Since
the ${^{QQ}\!G}$-function in the $QQ$ channel is explicitly known the
other ones can be unambiguously reconstructed \cite{1} and colour factors
trivially restored. The full NLO kernel has now the following form
\be
\label{def-str-NLO}
\mbox{\boldmath$V$}^{(1)}
= -
\mbox{\boldmath$\dot V$} \OO
\left(
\mbox{\boldmath$V$}^{(0)} + \ft{\beta_0}{2}\, \1
\right)-
\left[
\mbox{\boldmath$g$} \OO , \mbox{\boldmath$V$}^{(0)}
\right]_-
+ \mbox{\boldmath$G$}
+ \mbox{\boldmath$D$} .
\ee

%%%%%%%%%%%%%%%%%%%%%%%%%%%%%%%%%%%%%%%%%%%%%%%%%%%%%%%%%%%%%%%%%%%%%%%%
\section{Final reconstruction}
%%%%%%%%%%%%%%%%%%%%%%%%%%%%%%%%%%%%%%%%%%%%%%%%%%%%%%%%%%%%%%%%%%%%%%%%

The unknown remaining diagonal piece $\mbox{\boldmath$D$}$ can be
reconstructed by forming the forward limit to splitting functions, e.g.\
${^{QQ}P} (z) = {\rm LIM} {^{QQ}V} (x, y) \equiv \lim_{\xi \to 0}
{^{QQ}V} (\ft{z}{\xi},\ft{1}{\xi}) / |\xi|$. Comparing it with the known
two-loop DGLAP kernels one represents $\mbox{\boldmath$D$}$ as a
convolution of simple kernels whose non-forward counterparts are easy to
find. Since ${\rm LIM} \left\{ V_1 \OO V_2 \right\} = {\rm LIM} V_1 \OO
{\rm LIM} V_2$ restoration of $D$ from the forward case is simple and
one gets a complete $\mbox{\boldmath$V$}^{(1)}$ \cite{1}.

%%%%%%%%%%%%%%%%%%%%%%%%%%%%%%%%%%%%%%%%%%%%%%%%%%%%%%%%%%%%%%%%%%%%%%%%
\section*{References}
%%%%%%%%%%%%%%%%%%%%%%%%%%%%%%%%%%%%%%%%%%%%%%%%%%%%%%%%%%%%%%%%%%%%%%%%

\end{document}